\documentstyle[11pt,newpasp]{article}

\begin{document}

\title{Modeling Interacting Galaxies Using a Parallel Genetic Algorithm}
\author{Christian Theis and Stefan Harfst}
\affil{Institut f.\ Theoretische Physik und Astrophysik, Universit\"at Kiel,
24098 Kiel, Germany, email: theis@astrophysik.uni-kiel.de}

%-----------------------------------------------------------
%      abstract
%-----------------------------------------------------------

\begin{abstract}
    Modeling of interacting galaxies suffers from an extended parameter
space prohibiting traditional grid based search strategies. 
As an alternative approach a combination of a Genetic Algorithm (GA)
with fast restricted N-body simulations can be applied. A typical fit
takes about 3--6 CPU-hours on a PentiumII processor. Here we present
a parallel implementation of our GA which reduces the CPU-requirement
of a parameter determination to a few minutes on 100 nodes of a CRAY T3E.
\end{abstract}


%-----------------------------------------------------------
%      Introduction
%-----------------------------------------------------------

\vspace*{-1cm}

\section{Introduction}

   A major problem in modeling of encounters of galaxies is the extended
parameter space which is composed by the orbital and the structural
parameters of the interacting galaxies. Traditional grid-based fitting
strategies suffer from very large CPU-requirements. E.g.\ for
a restricted 7-dimensional parameter space (an encounter of 
a disc with a point mass) and a resolution of only 5 values per dimension, 
one needs 78125 models, or about 26 years of integration time
on a GRAPE3af special purpose computer for a ''complete'' grid. 
More systematic search strategies like gradient methods depend strongly
on the initial conditions, which makes them prone to
trapping in local optima. An efficient alternative approach are 
evolutionary methods and especially {\it genetic algorithms} (Holland 1975, 
Goldberg 1989, Charbonneau 1995). In combination with fast 
(but not self-consistent) {\it restricted 
N-body}-codes (Toomre \& Toomre 1972)
they allow for an efficient search in parameter space which can
be used for both, an automatic search of interaction parameters
(provided sufficiently accurate data are available) and/or a uniqueness
test of a preferred parameter combination (Wahde 1998, Theis 1999).

The basic idea of GAs is to apply an evolutionary mechanism including 
'sexual' reproduction operating on a population which represents a group of
different parameter sets. All members are 
characterized by their fitness which quantifies the correspondance between
simulations and observations. In order to determine the {\it 'parents'} 
two individuals are selected according to their fitness.
These parents represent two points in parameter space.
The corresponding parameter set is treated like a {\it 'chromosome'},
i.e.\ it is subject to a {\it cross-over} and a {\it mutation} operation
resulting in a new individual which is a member of the next generation.
Such a breeding is repeated until the next generation has been formed. 
Finally, the whole process of sexual reproduction is repeated iteratively
until the population confines one or several regions of sufficiently high 
fitness in parameter space.

   Here we present a parallel implementation of our genetic algorithm.

\section{Parallelization of the genetic algorithm}

 We parallelized our code on the level of the GA by applying a 
'master-slave' technique using {\it message passing} (MPI). 
On the master processor all the reproduction
operations are performed. It sends the individual parameter sets to the
slave processors which do the N-body simulations and determine the
corresponding fitness values (Fig.\ 1 left).
The speed-up, i.e.\ the ratio of CPU-times $t_1$ and $t_N$ used by
a calculation with one or $N$ processors, respectively,
reaches values between 30 and 60 for 100 processors (Fig.\ 1 right). 
The deviation from the optimal gain is mainly caused by the
spread of CPU-times required for the individual simulations. 
It becomes more important with decreasing population size. 
By parallelization the whole GA-fit 
using 10000 models reduces to less than 3 CPU minutes on a CRAY T3E.
Thus, an 'interactive' analysis becomes possible.

\begin{figure*}
   \includegraphics{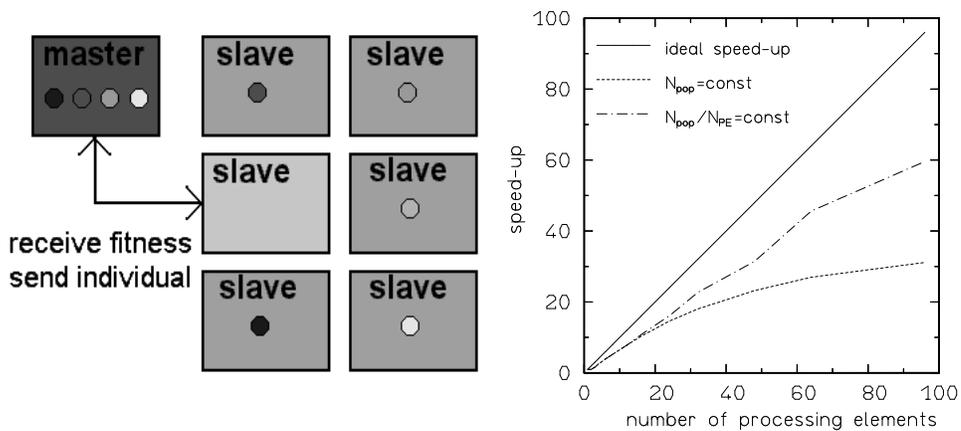}
   \vspace{6cm}
   \caption{ {\it left:} parallelization scheme; 
             {\it right:} speed-up vs.\ number $N_{\rm PE}$ of processors.
             Note the better performance, if the workload per processor
             ($N_{\rm pop} / N_{\rm PE}$) is kept constant ($N_{\rm pop}$ is
             the population size of 192 here). }
\end{figure*}

\acknowledgments
The authors are grateful to Paul Charbonneau and 
Barry Knapp for providing their (serial) genetic algorithm {\sc pikaia}.
C.T. thanks for the support by the organizers of the 15th IAP 
meeting.

\end{document}